\documentclass[aps,prx,reprint,floatfix,superscriptaddress,nofootinbib]{revtex4-2}
\usepackage{CJK}
\usepackage[utf8]{inputenc}
\usepackage{xcolor}
\usepackage{hyperref}
\usepackage{siunitx}
\usepackage{braket,amsmath,amssymb,mathtools,amsthm}
\usepackage[linesnumbered,ruled,vlined]{algorithm2e}

\newtheorem{theorem}{Theorem}

\newtheorem{definition}[theorem]{Definition}

\begin{document}
\begin{CJK*}{GB}{} 
\title{Modelling and Simulating the Noisy Behaviour of Near-term Quantum Computers}
\author{Konstantinos Georgopoulos}
\email{k.georgopoulos2@newcastle.ac.uk}
\affiliation{School of Computing, Newcastle University, Newcastle-upon-Tyne, NE4 5TG, United Kingdom}
\author{Clive Emary}
\affiliation{Joint Quantum Centre Durham-Newcastle, School of Mathematics, Statistics and Physics, Newcastle University, Newcastle-upon-Tyne, NE1 7RU, United Kingdom}
\author{Paolo Zuliani}
\affiliation{School of Computing, Newcastle University, Newcastle-upon-Tyne, NE4 5TG, United Kingdom}
\date{\today}

\begin{abstract}
    Noise dominates every aspect of near-term quantum computers, rendering it exceedingly difficult to carry out even small computations. In this paper we are concerned with the modelling of noise in Noisy Intermediate-Scale Quantum (NISQ) computers. We focus on three error groups that represent the main sources of noise during a computation and present quantum channels that model each source. We engineer a noise model that combines all three noise channels and simulates the evolution of the quantum computer using its calibrated error rates. We run various experiments of our model, showcasing its behaviour compared to other noise models and an IBM quantum computer. We find that our model provides a better approximation of the quantum computer's behaviour than the other models. Following this, we use a genetic algorithm to optimize the parameters used by our noise model, bringing the behaviour of the model even closer to the quantum computer. Finally, a comparison between the pre and postoptimization parameters reveals that, according to our model, certain operations can be more or less erroneous than the hardware-calibrated parameters show.
\end{abstract}

\maketitle
\end{CJK*}

\section{Introduction}\label{sec:intro}
Noise is a central obstacle in building large-scale quantum computers and executing long quantum computations. It is either due to infidelities of the quantum hardware (i.e., gates, measurement devices), or due to unwanted interactions with the environment (i.e., thermal, electromagnetic, gravitational decoherence) \cite{Brown-2004,Viola-1999,Bassi-2017,Pfister-2016}.

It has been proven that arbitrary long quantum computation is possible given constraints on the error rates and the error locality \cite{Gottesman-2010,Shor-1995,Knill-1998}. One method for mitigating quantum noise is through quantum error correction \cite{Shor-1995,Knill-1998,Calderbank-1996,DiVincenzo-1996,Knill-2005,Bacon-2006,Kitaev-2003,Dennis-2002,Bombin-2006,Duan-2010,Bombin-2010}, which relies on knowing what the most likely error sources are. It has been found that error correcting methods optimized for specific noise in a system can dramatically outperform generic ones \cite{Aliferis-2008,Tuckett-2018}. Thus, identifying, characterizing and simulating the noise in quantum computers is important and can lead to much more efficient calibration and error correction, which are necessary for large-scale quantum computing \cite{Martinis-2015}.

The most common error model is a depolarizing Pauli channel. Effectively, a Pauli operator is chosen to be applied at operations that have a probability to produce an erroneous result \cite{Knill-2005,Dennis-2002,Cross-2009}. This unital channel is generally a good approximation of most error processes that lead to a maximally mixed noisy state. Alternative depolarizing channels can use Clifford operations to effectively approximate quantum errors \cite{Gutierrez-2013}.

A large source of error comes from non-unital interactions between the quantum system and the environment. Noise of this type causes decoherence during the computation in various forms. The most common one is thermal relaxation/excitation, which plays a central role in our model. The non-unital nature of such quantum noise makes it difficult to simulate it with Pauli or Clifford operations, leading to a more complicated approach. Other forms of decoherence, like electromagnetic or gravitational, are more complex and thus not considered in this paper.

Within this work we are concerned with three main sources of error in a quantum computer: (i) gate infidelities, (ii) state preparation and measurement (SPAM) errors and (iii) thermal decoherence and dephasing of the physical qubits. Each noise source adheres to different aspects of the hardware and of the interaction between the system and its environment and is modelled as a quantum channel. We then proceed to combine these three quantum channels into a single, architecture aware noise model and we compare it against state-of-the-art models on quantum walk circuits implemented on an IBM quantum computer. The analysis shows that our unified model offers more accurate approximations of an IBM quantum computer's evolution and significant improvements in the accuracy of the noise simulations.

A very important role to the success of our model is played by the noise parameters utilized. These parameters represent various error rates and decoherence and dephasing times that result from calibrations of the actual quantum computer. We show that we can optimize a subset of the aforementioned parameters used by our noise model. This leads up to $84\%$ better approximation of the quantum computer's evolution. Another outcome of this analysis is the comparison between the hardware-calibrated parameters and the optimized parameters, providing evidence on what the computer's error rates are in the model, thus allowing further conclusions on the infidelities of the quantum hardware.

Finally, our noise model is not limited to modelling the IBMQ computers. Its architecture awareness and low-level circuit approach allows for it to be easily attached on any QASM-based implementation. For our noise model's implementation we make use of the IBM Qiskit development kit \cite{Qiskit,IBMQExp} in order to simulate and execute the quantum circuits.

Surprisingly, there are only a few works addressing the topic of modeling noise in quantum computers \cite{Harper-2020,Bogdanov-2013,Nachman-2020}. Notably in \cite{Lilly-2020-modeling}, the authors also attempt to generate a composite model for noisy quantum circuits by dividing the quantum circuit to subcircuits, according to desired characteristics. This decomposition allows for iterative adjustment of the models through minimization of the total variation distance between simulation and experimental results, until sufficient accuracy is obtained.

This paper is organized as follows. In Section \ref{sec:channels} we discuss the theoretical foundations of the noise channels by categorizing them into three error groups. Section \ref{sec:qcomb} shows how the error channels can be combined to a single, unified model for simulating noise, as well as a discussion about the noise parameters that are used by the models, before moving on to simulating the noisy evolution of a quantum system and comparing it to the real quantum computer in Section \ref{sec:simulations}. Section \ref{sec:optimize} showcases the optimization procedure for the noise parameters. Additional experiments with the optimized parameters are also carried out in this section. Finally, we present our conclusions and possible future work that could be undertaken in this area.

\section{Quantum Noise Channels}\label{sec:channels}
As mentioned in the introduction, we are concerned with three sources of error: (i) hardware infidelities in the form of depolarizing Pauli noise, (ii) state preparation and measurement (SPAM) errors and (iii) decoherence in the form of thermal relaxation and dephasing. In this section we discuss the three quantum channels we use to model each error.

\subsection{Error Group 1: Depolarizing Channel}\label{subsec:depol}
The first channel is also known as \textit{symmetric depolarizing channel}, a term which we will interchangeably use with \textit{gate infidelities}, or simply, \textit{depolarizing channel}. It essentially simulates the bit-flip and phase-flip errors due to gate infidelities within the circuit as a depolarizing channel \cite{nielsen_chuang_2010,wilde_2017,King-2003,Ji_2008}. We assume that an error of this group occurs with probability $p_{1}$, and we define the bit-flip and phase-flip errors through the Pauli $X$ and $Z$ operations. When both a bit- and phase-flip happen, the operation is defined through Pauli $Y$. All three types of Pauli errors have the same probability to occur. The depolarizing channel can be represented by the following operators
\begin{equation}
    \begin{aligned}
        K_{D_{0}} &= \sqrt{1-p_{1}}I, \\
        K_{D_{1}} &= \sqrt{\frac{p_{1}}{3}}X, \\
        K_{D_{2}} &= \sqrt{\frac{p_{1}}{3}}Z, \\
        K_{D_{3}} &= \sqrt{\frac{p_{1}}{3}}Y.
    \end{aligned}
\end{equation}

The effect of the depolarizing channel on a quantum system can be expressed via the operator-sum representation, as
\begin{gather*}
    \rho \mapsto \mathcal{D}(\rho) = \sum_{i=0}^{3}K_{D_{i}}\rho K_{D_{i}}^{\dagger}
\end{gather*}
where $\rho$ is the density matrix for a qubit. It is noteworthy that, as $K_{D_{i}}=K_{D_{i}}^{\dagger}$, we can do the relative replacement in the above representation.

\subsection{Error Group 2: State Preparation and Measurement (SPAM) Channel}\label{subsec:spam}
This channel is essentially a simple Pauli $X$ error, but we separate it from the above group as it refers to different aspects of the hardware and the computation. Thus, we can represent the SPAM quantum channel for the measurement errors by the following Kraus operators
\begin{equation}
    \begin{aligned}
        K_{M_{0}} &= \sqrt{1-p_{2}}I, \\
        K_{M_{1}} &= \sqrt{p_{2}}X
    \end{aligned}
\end{equation}
where $p_{2}$ is the probability that the measurement is incorrect.

The effect of the SPAM channel for measurement errors can be expressed through the density matrix, $\rho$, as
\begin{gather*}
    \rho \mapsto \mathcal{S}(\rho) = K_{M_{0}}\rho K_{M_{0}} + K_{M_{1}}\rho K_{M_{1}}.
\end{gather*}

In the case that state preparation takes place in the computation, the error channel (i.e., $\rho\mapsto\mathcal{S}^{\prime}(\rho)$) is of similar form to the measurement case, with the qubit failing to be prepared at the desired state, resulting to the inverted state by $X$ with probability $p_{2}^{\prime}$.

It is important to clarify the main idea behind separating the state preparation and measurement operations from the rest of the quantum circuit. On IBMQ, a state is prepared by injecting the standard initial state $\ket{0^{\otimes n}}$ to the register. Of course, quantum computations might start with a different initial state, which would require alternative operations for its preparation. Thus, we decide that the preparation of the quantum registers should not be part of the main execution of the quantum circuit that executes the algorithm. The reason for choosing measurement, on the other hand, as a separate quantum operation is self-explanatory. Finally, grouping them together in the same channel comes naturally as we deem both are modelled in the same way.

\subsection{Error Group 3: Thermal Decoherence and Dephasing Channel}\label{subsec:thermal}
The third error group refers to the physical qubits and their interaction with the environment. There are two aspects of noise within this error group: (i) the thermal decoherence (or relaxation) that occurs over time in the form of excitation/de-excitation and (ii) the dephasing of the qubits over time.

Thermal relaxation is a non-unital (i.e., irreversible) process that describes the thermalization of the qubit spins towards an equilibrium state at the temperature of their environment. This process involves the exchange of energy between the qubits and their environment, which drives the qubits either towards the ground state, $\ket{0}$ (de-excitation or reset to $\ket{0}$) or the excited state, $\ket{1}$ (excitation or reset to $\ket{1}$). On the other hand, dephasing refers to the ways in which coherence decays over time. It is a mechanism that describes the transition of a quantum system towards classical behaviour.

There already exists a function implementing this error group as a quantum channel within Qiskit\footnote{Thermal relaxation and dephasing channel in Qiskit: \url{https://qiskit.org/documentation/stubs/qiskit.providers.aer.noise.thermal_relaxation_error.html}} and details of the implementation can also be found in \cite{Blanco-2020}. The model takes into account:
\begin{itemize}
    \item the average execution time of each type of quantum gates $g$ implemented, denoted $T_{g}$;
    
    \item the time it takes for each qubit $q$ to relax and dephase, commonly denoted $T_{1}(q)$ and $T_{2}(q)$ respectively, where $q\in [0,n-1]$, where $n$ represents the number of qubits in the quantum computer.
\end{itemize}
In other words, $T_{1}(q)$ describes an evolution towards equilibrium as a perturbation orthogonal to the quantization axis ($x,y$-component of the Bloch vector) and $T_{2}(q)$ describes a slow perturbation along the quantization axis ($z$-component of the Bloch vector), or otherwise, the behaviour of the off diagonal elements over time for each qubit. These two times are related as $T_{2}(q) \leq 2T_{1}(q)$.

Considering the thermal relaxation and dephasing times $T_{1}(q)$ and $T_{2}(q)$, as well as the (known) gate execution times $T_{g}$, we can define the probability for each qubit $q$ to relax and dephase after a gate of type $g$ is applied to it as $p_{T_{1}}(q)=e^{-T_{g}/T_{1}(q)}$ and $p_{T_{2}}(q)=e^{-T_{g}/T_{2}(q)}$ respectively. We can then define the probability for a qubit to reset to an equilibrium state as $p_{\text{reset}}(q)=1-p_{T_{1}}(q)$.

Taking into account the thermal relaxation transition picture as described earlier (excitation and de-excitation), we can calculate the weight that dictates towards which of the two equilibrium states ($\ket{0}$ or $\ket{1}$) this noise (or reset error) drives each qubit, $q$, as \cite{Blanco-2020,Jin-esp}
\begin{equation}
    w_{e}(q) = \frac{1}{1+e^{2hf_{q}/k_{B}\Theta}}, \label{eq:esp}
\end{equation}
where $\Theta$ is the quantum processor's temperature, $h$ is Planck's constant, $k_{B}$ is Boltzmann's constant and $f_{q}$ is the frequency of the qubit.

This far we have taken into consideration the temperature of the quantum processor, $\Theta$. In general, according to IBMQ, the mixing chamber at the lowest part of the refrigerator brings the quantum processor and associated components down to a temperature $\Theta \approx 15$mK. As an example, considering the average frequency of the qubits within the IBM $15$-qubit Melbourne machine to be $\overline{f}_{q} \approx 4.9801 \times 10^{9}$Hz, we can calculate the average weight from equation \eqref{eq:esp} as $\overline{w}_{e} \approx 1.44532 \times 10^{-14}$. Thus, an excitation occurring with probability $p_{\text{reset}_{1}}=\overline{w}_{e}(1-p_{T_{1}})$ can be considered a rare event and can be omitted from our model. We then can effectively assume that the reset error takes the form of only reset to the ground state, $\ket{0}$, or in other words, that the device temperature is $\Theta = 0$. Thus, we can now refer to the thermal relaxation simply as relaxation or spontaneous emission. Important here is that our model assumes that the relaxation and dephasing noise occurs for each qubit in the system independently. Thus, for better presentation of the equations hence forth, we selectively omit the presence of the qubit identifier, $q$ (i.e., $p_{\text{reset}}$ instead of $p_{\text{reset}}(q)$).

If $T_{2}(q) \leq T_{1}(q)$ for every qubit, then the relaxation and dephasing noise can be expressed as a mixed reset and unital quantum channel \cite{Blanco-2020}. Assuming a device temperature $\Theta = 0$, we can identify the following forms of noise:
\begin{itemize}
    \item Dephasing. A phase-flip which occurs with probability $p_{Z}=(1-p_{\text{reset}})(1-p_{T_{2}}p_{T_{1}}^{-1})/2$.

    \item Identity. In this case, nothing happens to the qubit, or otherwise, the identity, $I$, occurs with probability $p_{I}=1-p_{Z}-p_{\text{reset}}$.
    
    \item Reset to $\ket{0}$. This represents a qubit's thermal decay, or a jump to the ground state. We can define the probability of a qubit to reset to the ground state as $p_{\text{reset}}=1-p_{T_{1}}$.
\end{itemize}

Having omitted the thermal excitation, we can represent the relaxation and dephasing channel with the following operators
\begin{equation}
    \begin{aligned}
        K_{I} &= \sqrt{p_{I}}I, \\
        K_{Z} &= \sqrt{p_{Z}}Z, \\
        K_{\text{reset}} &= \sqrt{p_{\text{reset}}} \ket{0}\bra{0}.
    \end{aligned}
\end{equation}
If we want to take the reset to $\ket{1}$ into account, the representation is similar and can be seen in Appendix \ref{ap:trm}.

Thus the effect of the relaxation channel when $T_{2}(q)\leq T_{1}(q)$ can be expressed as
\begin{gather*}
    \rho \mapsto \mathcal{N}(\rho) = \sum_{k\in \left\{ I,Z,\text{reset} \right\}}K_{k}\rho K_{k}^{\dagger}.
\end{gather*}

If $2T_{1}(q)\geq T_{2}(q) > T_{1}(q)$, the model implementation uses a Choi-matrix representation \cite{Choi-1972,Choi-1975}. The Choi matrix can be written as \cite{Blanco-2020}
\begin{equation}
    C = \begin{pmatrix} 1 & 0 & 0 & p_{T_{2}} \\ 0 & 0 & 0 & 0 \\ 0 & 0 & p_{\text{reset}} & 0 \\ p_{T_{2}} & 0 & 0 & 1-p_{\text{reset}} \end{pmatrix} \label{eq:choi}
\end{equation}
with the probabilities as defined above. 

The evolution of the density matrix $\rho$ with respect to the Choi matrix $C$ can be described as
\begin{equation*}
    \rho \mapsto \mathcal{N}(\rho) = \text{tr}_{1} \left[C (\rho^{T}\otimes I) \right],
\end{equation*}
where $\text{tr}_{1}$ is the trace over the main system in which the density matrix $\rho$ resides. The transition from Choi-matrix to operator-sum representation can be realised via the eigenvalues of the matrix, in the case they are non-negative and the matrix is Hermitian, or otherwise through singular value decomposition (see Appendix \ref{ap:trm}).

Finally, it is noteworthy that the thermal decoherence and dephasing model does not account for decoherence and dephasing effects during idle times of the qubits, as these effects are attributed mainly to the effects of electromagnetic interference and cross-talk between the qubits. What our model accounts for is decoherence and dephasing on idle qubits over time. More specifically, as the execution time of every quantum gate is known and used as a parameter within the thermal decoherence and dephasing model, the total execution time of the quantum circuit after every operation takes place is computed. Thus, when the probability of decoherence or dephasing of a qubit is calculated by the channel, the time passed from the start of the execution is taken into account.

\section{Unified Model for Quantum Noise}\label{sec:qcomb}
Following the individual definition of the three quantum noise channels, we now define the \textit{unified quantum noise model}.

\subsection{Quantum Noise Parameters}\label{subsec:noiseparams}
These parameters are used by the individual quantum noise channels and are usually given by calibration of the quantum computer. For the IBM quantum computers, the calibrated parameters are publicly available.

There are a few techniques used to calibrate the error rates and decoherence times of quantum computers, like cross-entropy benchmarking \cite{Boixo-2018-XEB,Aaronson-2019-XEB}, process tomography \cite{Zhou-2014,OBrien-2004,Mohseni-2008} or randomised benchmarking \cite{Onorati-2019,Magesan-2011,Lopez-2009,Knill-2008,Helsen-2019,Emerson-2005,Dankert-2009,Cross-2016}. Randomized benchmarking specifically is the most used and most prominent technique, with a few recent alterations like cycle benchmarking \cite{Erhard-2019-Cycle} or dihedral benchmarking \cite{Carignan-2015-Dihedral}.

For the first error group the noise parameters come in the form of operation error rates: they represent the probability $p_{1}$ that a gate, when applied in the quantum circuit, produces an erroneous outcome. Each individual type of gates implemented within the architecture (Pauli gates, Clifford gates, \verb|CNOT| etc) is associated with a specific error rate. Additionally, each type of gate has different error rates depending on the qubit(s) that they are applied on.

Similarly, the SPAM channel noise parameters are a selection of error rates that represent the probability that the preparation of the initial quantum state or the outcome of a measurement will be erroneous ($p_{2}^{\prime}$ and $p_{2}$ respectively). Each qubit in the system yields different error rates when prepared or measured.

Table \ref{table:errorates} shows the type of parameters with respect to each quantum noise channel, as well as how many parameters are associated with each error channel.

\begin{table}[!tb]
    \centering
    \begin{tabular}{c|c|c}
        \hline
        Error Group & Type of Parameter & Number of Parameters \\ \hline \hline
        Depolarizing & Error Rates $p_{1}$ & $r$ \\ 
        SPAM & Error Rates $p_{2}$ & $m+s$ \\ 
        Thermal Relax. & Times $T_{1}$ and $T_{2}$ & $2n$ \\ \hline
    \end{tabular}
    \caption{Type and number of parameters for each of the three error groups; $n$ is the number of qubits in the system, $m$ is the number of qubits that are measured, $s$ is the number of state preparations that occur and $r$ is the number of distinct types of gates implemented in the architecture, each considered once per qubit or pair of qubits.}
    \label{table:errorates}
\end{table}

\begin{figure*}[!tb]
    \begin{center}
        \includegraphics[width=17cm]{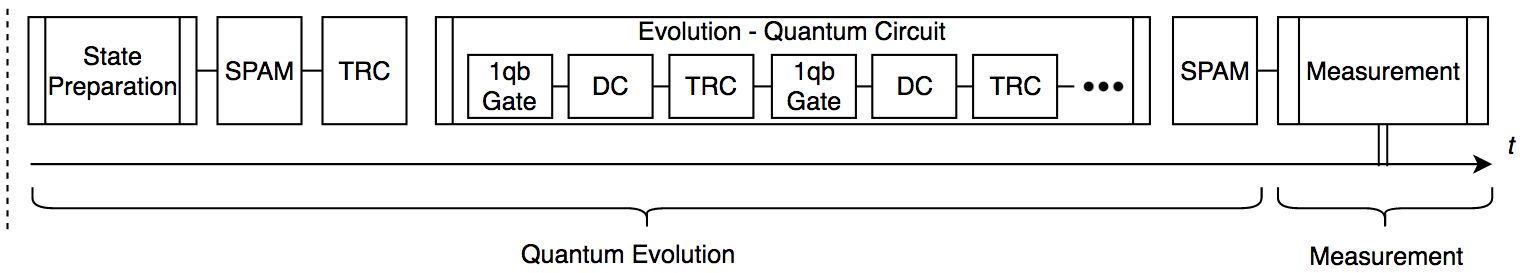}
    \end{center}
    \caption{The unified quantum noise model on a single-qubit circuit. The SPAM model is applied at the start, after the state preparation (if that occurs) and at the end before the measurement. The depolarizing channel (DC) is applied after every gate during the evolution of the circuit. The relaxation and dephasing channel (TRC) is applied after every gate and after the depolarizing channel.}
    \label{fig:combmodel}
\end{figure*}

\subsection{Constructing the Unified Model}\label{subsec:construct}
The main characteristic of our model is its \textit{architecture awareness}. The model takes into account the connectivity of the qubits within the architectural graph of the computer, as well as the specific properties of the qubits (i.e., decoherence time) and the gates (i.e., execution time, error rates) that participate in the system.

\paragraph{Depolarizing Channel.} A circuit executed directly on a quantum computer includes either single- or two-qubit gates. We can construct the depolarizing quantum channel according to the following rules:
\begin{enumerate}
    \item Single-qubit errors occur \textit{after} a single-qubit gate in compliance with the single-qubit error rates.
    
    \item Two-qubit errors occur \textit{after} a two-qubit gate according to the two-qubit error rates. Here, in the context of an architecture-aware model, the knowledge of the computer's qubit connectivity is encoded within the model.
\end{enumerate}

\paragraph{SPAM Channel.} State preparation errors take place \textit{after} the state preparation, if that occurs, and measurement errors occur \textit{before} measurement according to their respective error rates.

\paragraph{Thermal Relaxation and Dephasing Channel.} Finally, we can apply the relaxation and dephasing channel as a function on each individual qubit in the system. This function is implemented \textit{after} each gate is applied and occurs according to the relaxation and dephasing times of each qubit in the system, as well as the duration of each type of quantum gate within the system.

\paragraph{Unifying the Channels.} We have implemented our model in Qiskit, which can simulate the thermal decay and dephasing quantum channel. Additionally, this channel requires the average execution time of each type of gate. These times are assumed to remain static for each quantum computer.

Having a set of guidelines on the construction of the individual quantum channels, we can now easily create the \textit{unified quantum noise model} as the combination of the three noise channels. The application of every quantum channel is independent and their combination is simply computed by composing the error operators with the circuit gates. Assuming an arbitrary number $t$ of unitary, single-qubit quantum gates $U_{t}$, and an initial quantum state $\rho_{0}$, we can express the effect of the unified noise model on the evolution by the following operator
\begin{equation}
    \mathcal{V} = M \cdot \mathcal{S} \cdot \prod_{t}\Big( \mathcal{N} \cdot \mathcal{D} \cdot \mathcal{U}_{t} \Big) \cdot \mathcal{N} \cdot \mathcal{S}^{\prime} \cdot P (\rho_{0})
\end{equation}
where $\mathcal{U}_{t}(\rho)=U_{t}\rho U_{t}^{\dagger}$, $\mathcal{D}$, $\mathcal{S}$, $\mathcal{S}^{\prime}$ and $\mathcal{N}$ are the depolarizing, measurement, state preparation, and relaxation  and dephasing channels respectively, $M$ is a measurement superoperator, $P$ is a state preparation superoperator. This definition can be readily expanded to account for higher dimensional operators (e.g., for two-qubit operations). Figure \ref{fig:combmodel} visualizes the unified quantum noise model for the single-qubit case.

\begin{figure}[!t]
    \begin{tabular}{c}
          \includegraphics[width=7.8cm]{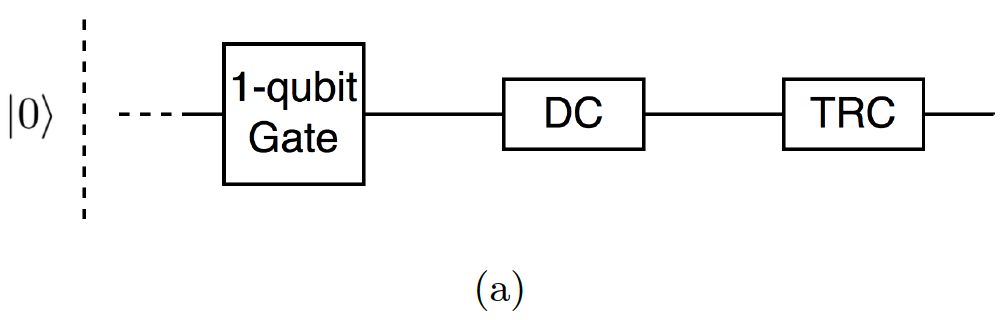} \\
          \includegraphics[width=8.4cm]{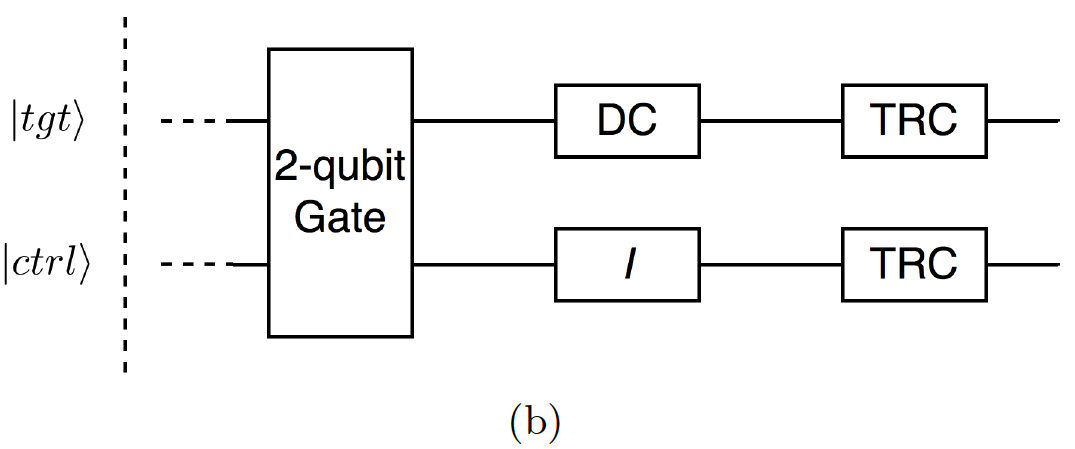}
    \end{tabular}
    \caption{(a) Application of depolarizing (DC) and relaxation and dephasing (TRC) channels on a single-qubit gate. (b) Application of the two channels on a two-qubit gate; $\ket{ctrl}$ and $\ket{tgt}$ are control and target qubits; the channels are applied independently on each qubit and \textit{I} is the identity, which is considered a virtual gate with zero execution time.}
    \label{fig:channelapp}
\end{figure}

Finally, it is noteworthy how the model treats single- and two-qubit gates differently when the depolarizing and relaxation and dephasing channels are applied. After each gate in the circuit, the two channels occur independently of each other and can be combined by composition. Figure \ref{fig:channelapp} visualises the effect of the channels on each type of gate. Specifically, in the two-qubit gate, we observe that only the target qubit is affected by the depolarizing channel. This happens as, within our model, the part of the operation that has a chance to go wrong is the ``state change''. In other words, the control qubit acts just as a driver of the quantum gate, and the gate has no effect on its state, either willingly or through the effects of noise.

\section{Simulating Noise in Quantum Computers}\label{sec:simulations}
For our experiments, the unified quantum noise model is implemented using Python and the circuits are executed using the IBMQ Qiskit simulators \cite{Qiskit} and the $15$-qubit Melbourne computer. Of course, this does not induce any difficulties in applying the model in different architectures that use QASM or QASM-type implementation for the low level quantum circuits. The code is available on GitHub \cite{GithubRepo}.

\subsection{Preliminary Methods}\label{subsec:prel}
For our experiments we use discrete-time and space quantum walks. The main reason for this choice is the predictable behaviour and the susceptibility of this algorithm to quantum noise. In addition, to test the performance of our model we need a metric to compare the results of the simulated noisy evolution and the execution on the quantum computer. For this purpose we use the Hellinger distance. Below we present a brief description of these two subjects.

\paragraph{Discrete-time Quantum Walks.} Quantum walks are unitary processes that describe the quantum mechanical analogue of a random walk on a graph or a lattice \cite{Aharonov-2001,KGeorgo-2020-rots,Kempe-2003}. They possess intrinsic properties that make them highly susceptible to quantum noise. First of all, discrete-time quantum walks exhibit modular behaviour \cite{KGeorgo-2020-rots,Reitzner-2011-mod}. This characteristic describes the modular relationship between the parity of the number of coin-flips of the walk, the initial state and the current position of the walker, a property that gets violated in a noisy environment \cite{KGeorgo-2020-rots}. For example, a walker initialized in an even state (e.g., $\ket{2}$), after an odd number of steps (i.e., $1$) will be found on an odd state (e.g., $\ket{1}$ or $\ket{3}$). The second property is that quantum walks propagate quadratically further than classical random walks \cite{Aharonov-2001,Szegedy}.

For the implementation of quantum walks, we will use a gate efficient approach that uses inverter gates, as shown in \cite{Douglas-2009-EffWalk}. In a previous work \cite{KGeorgo-2020-rots} we found that the number of gates in the circuit increases with the size of the state space, $N$, as $\mathcal{O}(\log^{2}{N})$. More details on the quantum walk circuit can be found in Appendix \ref{ap:qwcirc}.

\paragraph{Hellinger Distance (HD).} To compare the probability distributions of the noise model against the distributions generated by the quantum computer, we use the Hellinger distance \cite{Jin_2018}.
\begin{definition}[Hellinger distance]\label{def:hd}
    For probability distributions $P=\{p_{i}\}_{i\in[s]}$, $Q=\{q_{i}\}_{i\in[s]}$ supported on $[s]$, the Hellinger distance between them is defined as
    \begin{equation}
        h(P,Q) = \frac{1}{\sqrt{2}} \sqrt{ \sum_{i=1}^{k} \left( \sqrt{p_{i}} - \sqrt{q_{i}} \right)^{2} } \label{eq:hd}.
    \end{equation}
\end{definition}

The Hellinger distance is a metric satisfying the triangle inequality. It takes values between $0$ and $1$ (i.e. $h(P,Q)\in[0,1]$) with $0$ meaning that the two distributions are equal. Additionally, it is easy to compute, easy to read and it does not depend on the probability distributions having the same support. The last property is particularly useful since in many ideal output distribution of quantum circuits the probability mass is concentrated on a few states.

\paragraph{Model Parameters.} As described in Section \ref{subsec:noiseparams}, there are several parameters corresponding to each of the error groups we simulate. The architecture awareness of the model takes into account the individual error rates and decoherence times for each quibit separately, as well as for each pair of qubits through the connectivity of the architecture. Table \ref{table:params} showcases an example of the average for each category of error rates used in our model, calculated on the date of the experiments.

\bgroup
\def\arraystretch{1.2}%
\setlength\tabcolsep{0.12cm}
\begin{table}[!t]
    \centering
    \begin{tabular}{c|c|c|c|c}
        \hline
        $1$qb Errors & $2$qb Errors & Meas. Errors & $T_{1}$ $(\SI{}{\micro\second})$ & $T_{2}$ $(\SI{}{\micro\second})$ \\ \hline \hline
        $11.68\times 10^{-4}$ & $3.17\times 10^{-2}$ & $7.61 \times 10^{-2}$ & $56.15$ & $56.01$ \\ \hline
    \end{tabular}
    \caption{Average noise parameters for all the qubits of the IBMQ $15$-qubit Melbourne machine used on the date of the experiments. $1$qb Errors are the single-qubit gate errors, $2$qb Errors the two-qubit gate errors, Meas. Errors the readout errors and $T_{1}$ and $T_{2}$ the average relaxation and dephasing times of all $15$ qubits.}
    \label{table:params}
\end{table}
\egroup

\subsection{Experiments and Results with the Calibrated Parameters}\label{subsec:experiments}
For our experiments we will run one step of the quantum walk (i.e., one coin-flip), as previous work shows that this duration is satisfactory for errors to take place and the behaviour of the quantum walk to evolve in a predictable manner \cite{KGeorgo-2020-rots}. In general, we use as initial state for our quantum walks the state $\ket{0}$, which means we do not need to deal with state preparation errors.

Alongside our unified quantum noise model (UNM), we evaluate four additional noise models:
\begin{itemize}
    \item QiskitCM: a combination of a readout error, a depolarizing error and a relaxation and dephasing error implemented within Qiskit\footnote{More concrete description in Qiskit documentation: \url{https://qiskit.org/documentation/apidoc/aer_noise.html}},
    
    \item DSPAM: a simpler version of the UNM that includes the depolarizing model for the gate infidelities and the SPAM model for the measurement errors,
    
    \item TRM: a standalone relaxation and dephasing model implemented in Qiskit that follows the main principles of Error Group $3$ (Section \ref{subsec:thermal}) and
    
    \item SDM: a simple depolarizing model that is not architecture-aware.
\end{itemize}
This allows for a clearer comparison of the model performance on approximating the noisy behaviour of the computer. The QiskitCM model is clearly the more complex of the IBMQ noise simulators and, shares similarities with the UNM on the way it computes the error. On the other hand, QiskitCM does not take into account the noise parameters for each qubit separately, but calculates and utilizes their averages, a fact that is reflected through a larger deviation from the quantum computer distribution than the UNM (see Table \ref{table:decoptresults}). The DSPAM and TRM models are, essentially, a separate and simple implementation of Error Groups $1$ and $2$, respectively. Finally, the SDM model is just a simple depolarizing model that is completely architecture-aware, i.e., it does not take into account the connectivity of the qubits within the QPU, but instead, computes the noise through a simple probabilistic application of Pauli errors during the computation.

\begin{figure*}[!t]
    \begin{tabular}{cc}
          \includegraphics[width=7.5cm]{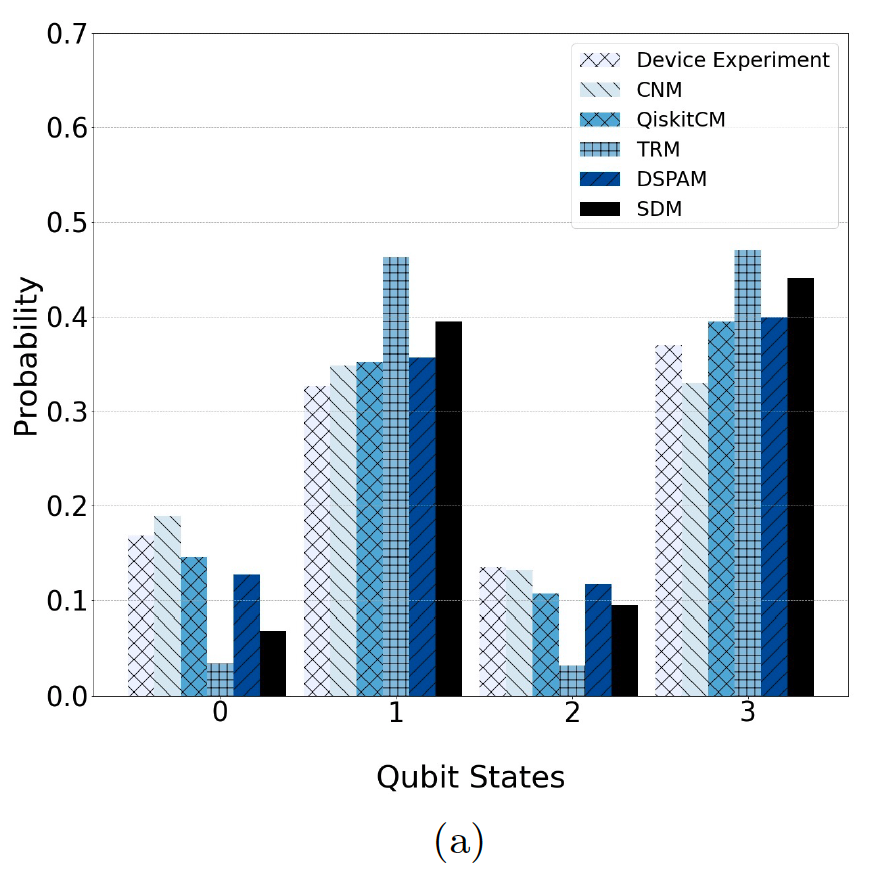} & \hspace{3em} \includegraphics[width=7.5cm]{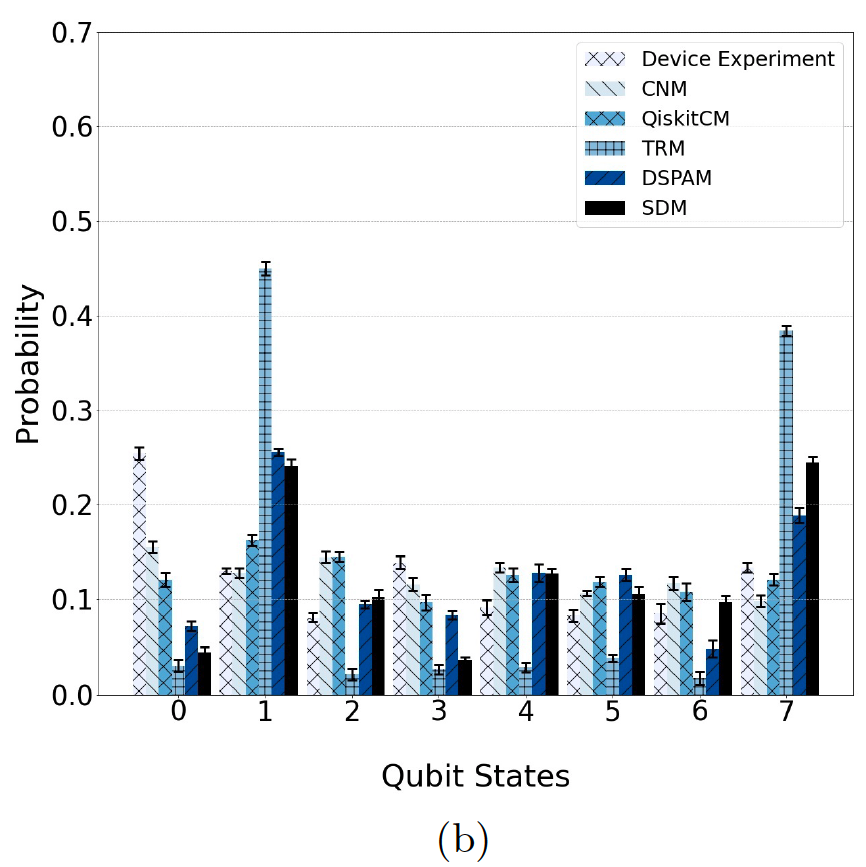}
    \end{tabular}
    \caption{(Color online) Comparison between the probability distributions of a quantum walk on (a) a two-qubit system and (b) a three-qubit system, simulated with the UNM (light-coloured vertically lined bar), QiskitCM (dark-coloured crossed bar), TRM (tiled bar), DSPAM (dark-coloured vertically lined bar), SDM (solid bar) and run on the actual quantum computer (light-coloured crossed bar). The quantum walk propagates for one coin-flip; error bars with $95\%$ confidence intervals are shown for the three-qubit system; for the two-qubit system they are smaller than $10^{-3}$, hence are not displayed.}
    \label{fig:2and3qQW}
\end{figure*}

Our experimental methodology consists of $100{,}000$ runs of the quantum walk, with the configurations described above, on the quantum computer and as a simulation with each of the noise models introduced above. We are interested in quantifying how close each model's evolution is to the quantum computer. Thus, we compute the Hellinger distance (HD) between the distribution of each simulated noise model and the computer. Figure \ref{fig:2and3qQW}(a) shows that on a two-qubit system, the unified quantum noise model provides a better approximation of the quantum computer's distribution than the other noise models. A numerical comparison of this result is shown in Table \ref{table:hd} (line $N=4$).

\bgroup
\def\arraystretch{1.2}%
\setlength\tabcolsep{0.325cm}
\begin{table*}[!t]
    \centering
    \begin{tabular}{c|c|c|c|c|c|c|c|c}
        \hline
        No. States ($N$) & No. Qubits ($\log{N}$) & UNM & QiskitCM & DSPAM & TRM & SDM & Ideal & Uniform \\ \hline \hline
        $4$ & $2$ & $0.033$ & $0.040$ & $0.049$ & $0.229$ & $0.126$ & $0.264$ & $0.218$ \\
        $8$ & $3$ & $0.127$ & $0.152$ & $0.224$ & $0.438$ & $0.280$ & $0.771$ & $0.186$ \\
        $16$ & $4$ & $0.224$ & $0.262$ & $0.369$ & $0.476$ & $0.329$ & $0.834$ & $0.150$ \\
        $32$ & $5$ & $0.393$ & $0.421$ & $0.482$ & $0.505$ & $0.467$ & $0.862$ & $0.434$ \\
        $64$ & $6$ & $0.457$ & $0.509$ & $0.525$ & $0.587$ & $0.576$ & $0.891$ & $0.579$ \\ \hline
    \end{tabular}
    \caption{Hellinger distance between the probability distributions of the quantum computer and the various noise models, as well as the ideal and uniform distributions. For ease of presentation, the acronyms are ascribed as UNM: unified noise model, QiskitCM: the Qiskit composite model, DSPAM: depolarizing and SPAM, TRM: relaxation and dephasing model, SDM: simple depolarizing model, Ideal: the theoretical distribution from an ideal (noise-free) quantum walk,  Uniform: uniform distribution, for the maximum-entropy guess.}
    \label{table:hd}
\end{table*}
\egroup

The next experiment repeats the above methodology for a quantum walk on three qubits. The respective results are shown in Figure \ref{fig:2and3qQW}(b). Again, the results showcase the superiority of the unified quantum noise model to the rest, with the smallest HD of $0.12749$ from the computer. This value, and indeed the distances of all the models from the computer, are much higher than the corresponding figures for smaller, two qubits quantum walk. This tells us that the models perform worse in approximating the noisy evolution of bigger quantum circuits.

Additional results from further experiments are shown in Table \ref{table:hd}. In general, we can see that the distance from the quantum computer's distribution is increasing with the number of qubits in the system. This is true for all the models. It is easy to realise that our UNM performs best, followed closely by QiskitCM.

As a final remark, it becomes apparent by the experimental results that the quantum computer, for quantum walks of size bigger than $N=8$, produces probability distributions that are closer to the uniform distribution due to excessive noise. Nevertheless, these results are driven by the effects and intensity of the noise within the QPU, and thus, are included in this analysis.

\subsection{Unified Noise Model vs Gate Set Tomography}\label{subsec:unmvsgst}
One of the prominent protocols for characterizing quantum operations is \textit{gate set tomography} (GST) \cite{Merkel-GST}. GST has been used in a large number of experiments \cite{Kim-2015,Deho-2016,Blume-2017,Ware-2021,Proctor-2020,Hong-2020,Manoj-2020} and implemented in open-source software \cite{pyGSTi,Nielsen-2020-pyGSTi}. The basic aim of GST is to characterize quantum operations performed by hardware. GST allows one to estimate the performance for a system with a small number of qubits. Additionally, it reconstructs or estimates not a single logic operation, but an entire set of logic operations (hence, gate set).

The characteristics of GST give rise to a meaningful comparison to our UNM. The unified model aims to reconstruct (simulate) the entire quantum evolution of a circuit. It follows the quantum circuit execution at runtime and simulates the effects of noise on three levels, gate infidelities, state preparation and measurement and decoherence and dephasing of the qubits. On the other hand, the GST is aimed to predictive characterization of the quantum gates of the circuit within the QPU, i.e., how the logic operations affect the qubits they act upon. The quantum gates need to be specified before the GST reconstructs the gate-driven evolution.

Furthermore, the GST protocol works well with \textit{small} quantum systems \cite{Nielsen-2020-pyGSTi,Nielsen-2020} whereas our UNM aims to approximate the noisy evolution within a QPU irrespective of the size of the quantum system. Thus, whereas GST works well for two- or three-qubit systems, the UNM is designed to scale with the size of the quantum circuit and the number of qubits. Of course, there is an upper bound on the scaling capabilities of the UNM tied to the increasing difficulty of classical machines to simulate increasing number of qubits.

Finally, one characteristic of the GST protocol is that it is calibration-free \cite{Nielsen-2020}. When GST reconstructs a model of a quantum system, it does not depend on any prior description of the measurements used or the states that can be prepared. The UNM, as is evident from the analysis above, depends on a set of quantum noise parameters, which reflect the levels of noise within the QPU during the execution of the quantum circuit. Thus, calibration of these noise parameters (or error rates) is essential for the success of the unified noise model, a reason that leads to the development of the noise parameter optimization technique showcased in the following section.

\subsection{Noise Modelling in Quantum Error Correction}\label{subsec:qec}
One of the most prominent fields in the development of quantum computing is \textit{quantum error correction} (QEC). Within QEC there is a large amount of literature and research along the lines of error analysis with NISQ systems. Similarly to our research, noise within a quantum computer is categorized in coherent systematic gate errors, environmental decoherence and models of loss, leakage, measurement and initialization errors \cite{Roffe-2019-QECintro,Devitt-2013-QECbeginners}.

Systematic noise contains any errors caused by faults of the quantum gates themselves, much like the gate infidelities within the UNM. Environmental decoherence tries to highlight how QEC relates to environmental effects. An elegant model for characterizing decoherence on open quantum systems is the Lindblad formalism \cite{Barnett-1993,nielsen_chuang_2010,Daffer-2003}, accompanied with several assumptions that may not hold in some cases \cite{Hope-2000,Martinis-2005,Astafiev-2004,Ahn-2002}. Particularly in superconducting systems where cross-talk and fluctuating charges can cause decoherence, the need arises for more specific decoherence models. One way to construct such models is via more general mappings \cite{Devitt-2013-QECbeginners}, or alternatively, a combination of models like the UNM presented in this work.

A recent paper \cite{Weber-2021-QECmodeling} discusses a structure for QEC which relies, amongst others, on evaluating noise modeling techniques or combinations of them. Our UNM is a perfect fit for such an approach by combining the major sources of error that play a significant role in quantum error mitigation.

Finally, it is evident that more complex or expanded quantum channels have the ability to better recreate the quantum noise before any type of QEC is applied \cite{Guttierez-2015-QEC}. Hence, incorporation of further types of systematic noise, like Clifford errors, or environmental decoherence, like electromagnetic noise, to the UNM, can lead to an even more accurate model for quantum noise.

\section{Optimizing the Quantum Noise Parameters}\label{sec:optimize}
Up to this point, the noise parameters used in our model are the ones calibrated from the computer itself. As evident by the experiments in Section \ref{sec:simulations}, that provides us with approximations that deviate from the quantum computer's evolution. In this section we implement a classical methodology that allows us to optimize the aforementioned noise parameters for the UNM and mimic the evolution of the quantum computer much more closely. Notably, such a procedure is possible for our unified noise model as it is easy to alter and feed the noise parameters to the model before a simulation. This is not possible, for example, when using the QiskitCM model as it automatically draws the hardware-calibrated parameters from the IBM computer itself and then constructs the model, a procedure we have no control over. This is indeed a limitation of the software implementation and not of the mathematical model itself.

As shown in Section \ref{subsec:noiseparams} there is a large number of parameters associated with each error group and their number grows with the size of the state space of the walk. Within the scope of this research we are working with the parameters associated with the depolarizing Pauli and SPAM models, i.e., error groups $1$ and $2$. The size of this set of parameters can be calculated as $r+m+s$, where $m$ is the number of qubits in the system that are measured, $s$ the number of qubits that undergo state preparation and $r$ is the number of different types of gates that are included in the circuit relative to the architecture of the quantum computer.

The decision to exclude the relaxation and dephasing parameters, $T_{1}(q)$ and $T_{2}(q)$, is taken for two main reasons. First of all, these parameters have to be taken into account per qubit, (hence the $(q)$). This means that with larger workspaces, the number of parameters for optimization grows very fast, rendering parameter optimization exceedingly taxing. This also means that, since there are more parameters to optimize, for the same number of generations, a smaller space of the possible optimal parameters will be explored, lowering the performance of the genetic algorithm. Secondly, the runtime of the parameter optimization becomes increasingly larger. (A further analysis regarding parameter optimization including the decoherence rates is given in Appendix \ref{ap:decopt}. The results show that the increase in efficiency of noise simulations is not enough to justify the associated increase in computational resources.) Throughout our work we find that the relaxation and dephasing of the qubits are parameters tied closely with the physical implementation of the qubits themselves within each quantum computer. Thus, we believe that an optimization of those parameters would prove more valuable when implemented on a qubit engineering modelling level.

In our quantum walks experiments, we are concerned with an implementation that includes Hadamard, inverter and \verb|CNOT| gates. Important here is that every gate will be considered for the parameter count only once per qubit or pair of qubits, no matter how many times it is used in the circuit. Thus, the number of parameters that need optimizing is $(1 + r_{c} + r_{t})+m+s$, where $m$ is the number of qubits measured, $s$ the number of state preparations, $r_{c}$ is the number of inverter gates, $r_{t}$ the number of \verb|CNOT| gates and $+1$ for the Hadamard gate. It is noteworthy that, due to tiny differences in the error rates of single-qubit gates, we can omit the differentiation between Hadamard and inverter gates without needing to optimize both types. Thus, the number of parameters can be calculated as $r_{s} + r_{t} + m + s$, with $r_{s}$ being the single-qubit gates.

\subsection{Parameter Optimization}\label{subsec:paramopt}
In order to obtain a set of better parameters we use a method based on \textit{genetic algorithms} (GA) \cite{GA-2998}. This method relies on iterative generations of new parameters, simulations using said new parameters and comparison of the simulated results with the quantum computer's distribution. In each iteration, the parameters that bring the simulated evolution closer to the quantum computer are kept. 

In order to keep the execution time small and the results presentable, we will again use a quantum walk with a small state space of $N=4$ and a three-qubit system for its execution. Here we need three qubits as one is necessary for the quantum coin. The coin is never measured, meaning the results of its error rate's optimization will not be directly visible, but through the overall effects on the computation. 

\begin{figure}[!tb]
    \begin{center}
        \includegraphics[width=8cm]{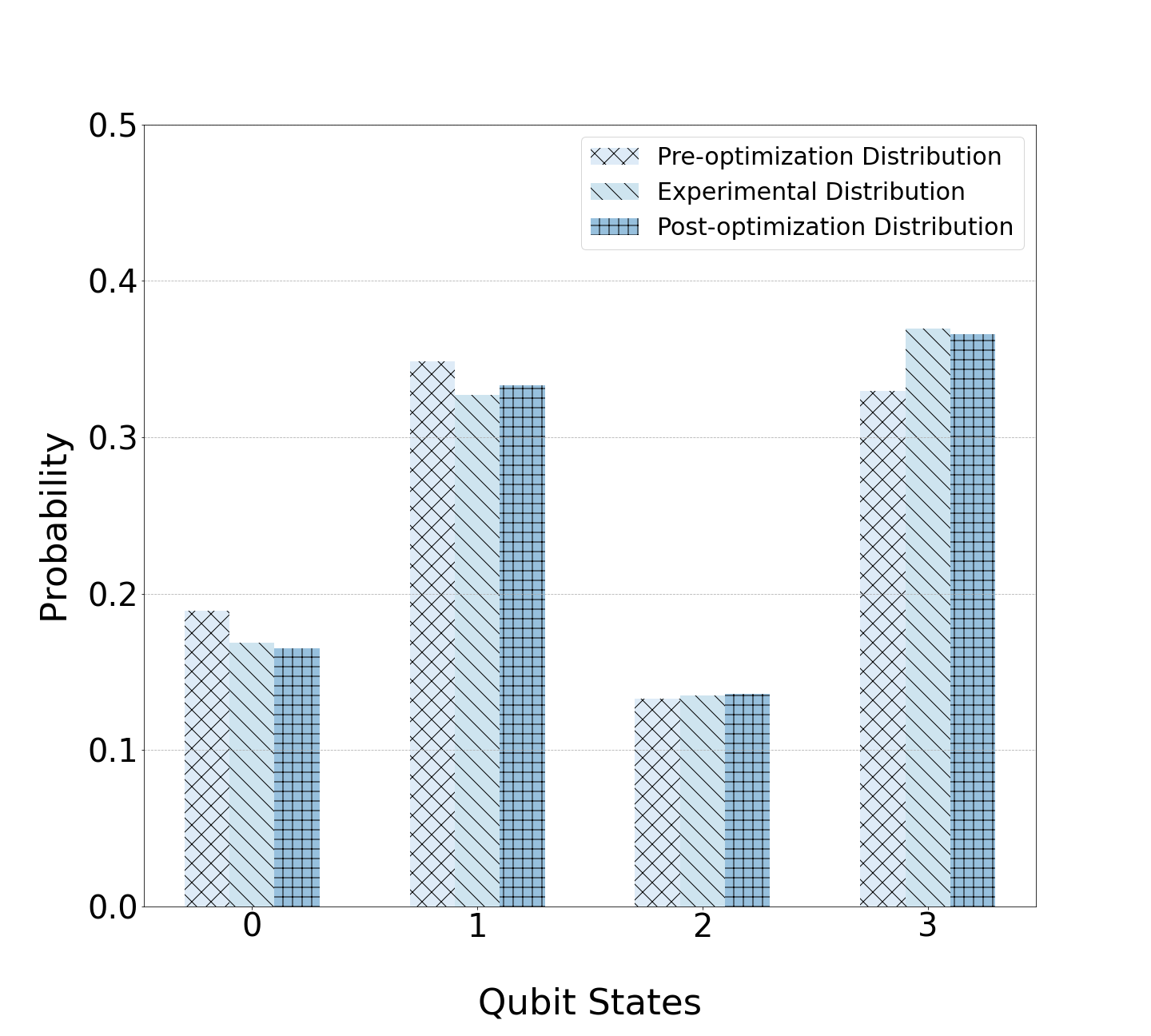}
    \end{center}
    \caption[]{(Color online) Probability distribution of pre and postoptimization (crossed and vertically lined bars respectively) after a single optimization routine compared with the quantum computer distribution (tiled bar). The optimized set produces a distribution that is $84.85\%$ closer to the quantum computer's.}
    \label{fig:2qpost}
\end{figure}

A comparison between pre and postoptimization for the two-qubit quantum walk after a single optimization routine is shown in Figure \ref{fig:2qpost}. For this task, we allow for $50$ generations of the genetic algorithm. The number of parameters that undergo optimization for the experiment on the IBMQ $15$-qubit Melbourne machine is $9$: $r_{s}=4$ single-qubit gate error rates, one for each of the four qubits in the system, $r_{t}=3$ two-qubit gates according to the architecture of the computer, one for each pair of connected qubits, and $m=2$ measurements at the end of the computation. As the computation is initialized at state $\ket{0}^{\otimes (\log{N})}$, we do not account for state preparation of the qubits ($s=0$). The HD between the postoptimization simulation of the quantum circuit and the quantum computer has decreased from $\sim 0.033$ to $\sim 0.005$, an approximately $84.85\%$ improvement.

\begin{figure}[!tb]
    \begin{center}
        \includegraphics[width=8cm]{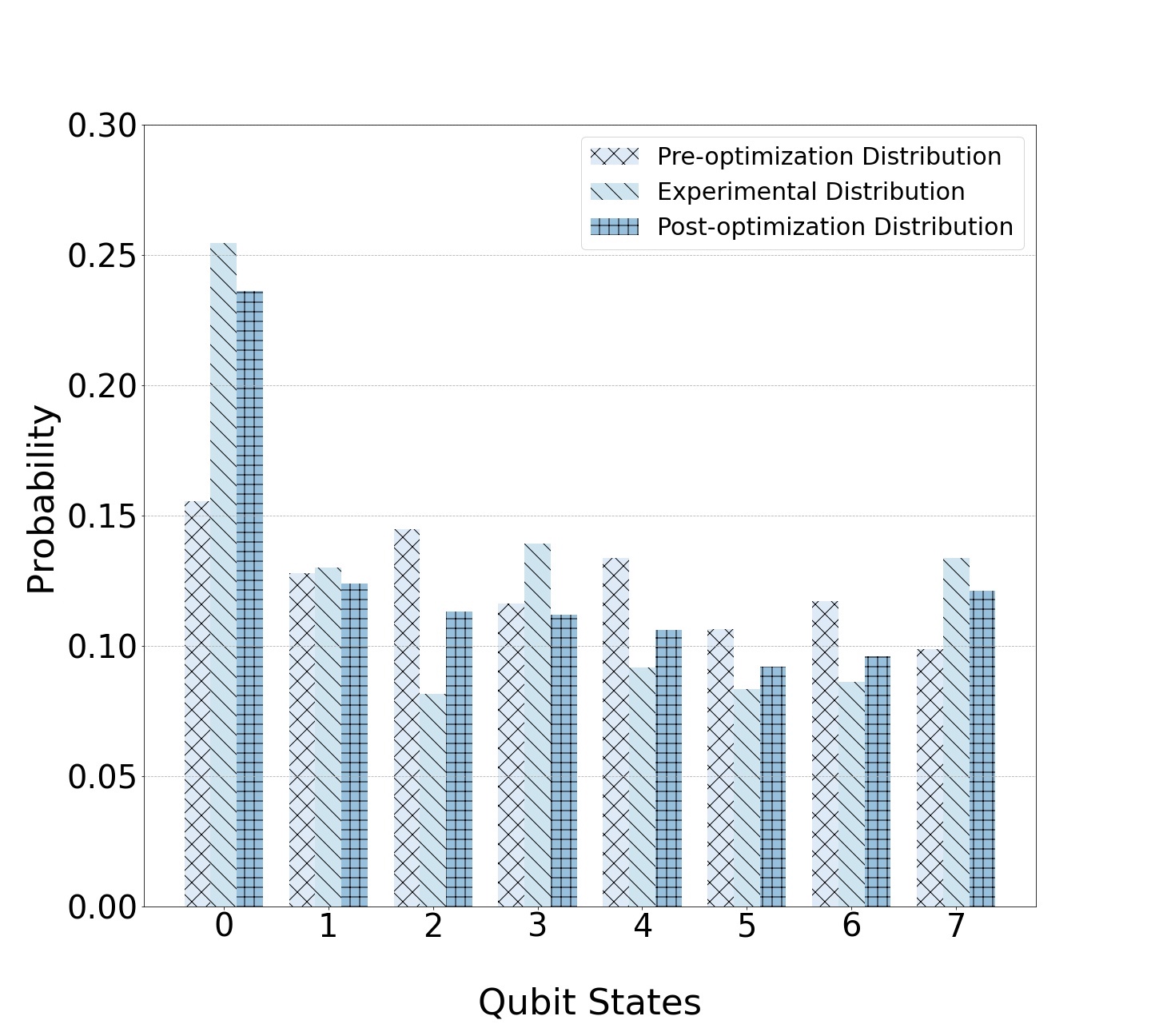}
    \end{center}
    \caption[]{(Color online) Probability distribution of pre and postoptimization (crossed and vertically lined bars respectively) after a single optimization routine compared with the quantum computer distribution (tiled bar). The optimized set produces a distribution that is $57.48\%$ closer to the quantum computer's.}
    \label{fig:3qpost}
\end{figure}

We can use the same methodology for a quantum walk on the Melbourne quantum computer with a state space of $N=8$. In this case, due to the nature of the implementation of the circuit, we will require an ancilla register \cite{KGeorgo-2020-rots}, which takes the number of qubits in the system to six. The number of parameters that require optimization is $14$: $r_{s}=6$ single-qubit rates, $r_{t}=5$ two-qubit rates, $m=3$ qubits measured -- the state space of the quantum walk -- and $s=0$ as we initialize on $\ket{0}$. For consistency, the GA is evolved for $50$ generations. Additionally, we calculate the runtime of the optimization routine on the classical computer. The results after a single optimization routine are shown in Figure \ref{fig:3qpost}. The HD between the postoptimization simulation of the quantum circuit and the quantum computer has decreased from $\sim 0.127$ to $\sim 0.054$, an approximately $57\%$ improvement.

Similar results are given for larger state spaces of the quantum walk. Table \ref{table:optresults}(a) presents the averaged results from three optimization routines, i.e., three runs of the genetic algorithm routine for $50$ generations each. Due to the need of ancillary qubits in the computation, the number of parameters that need optimization becomes large quickly. This means that the GA routine becomes slower with every qubit added in the state-space register.

\begin{table*}[!t]
    \begin{tabular}{c}
        \bgroup
        \def\arraystretch{1.2}%
        \setlength\tabcolsep{0.32cm}
            \centering
            \begin{tabular}{c|c|c|c|c|c}
                \hline
                No. States ($N$) & Size of Workspace & HD (Pre) & HD (Post) $\pm$ s.d. & $\%$ Distance & CPU Runtime ($\times 10^{3}$ sec) \\ \hline \hline
                $4$ & $4$ & $0.033$ & $0.005\pm 0.001$ & $84.85\downarrow$ & $6.5$\\
                $8$ & $6$ & $0.127$ & $0.054\pm 0.003$ & $57.48\downarrow$ & $9.5$ \\
                $16$ & $8$ & $0.224$ & $0.152\pm 0.006$ & $32.14\downarrow$ & $12.1$ \\
                $32$ & $10$ & $0.393$ & $0.301\pm 0.025$ & $23.41\downarrow$ & $46.7$ \\
                $64$ & $12$ & $0.457$ & $0.377\pm 0.016$ & $17.51\downarrow$ & $93.5$ \\ \hline
            \end{tabular}
        \egroup \\ [6pt] (a) Optimization with $50$ generations. \\ [6pt]
        \bgroup
        \def\arraystretch{1.2}%
        \setlength\tabcolsep{0.36cm}
            \centering
            \begin{tabular}{c|c|c|c|c|c}
                \hline
                No. States ($N$) & Size of Workspace & HD (Pre) & HD (Post) & $\%$ Distance & CPU Runtime ($\times 10^{3}$ sec) \\ \hline \hline
                $4$ & $4$ & $0.033$ & $0.003\pm 0.001$ & $88.26\downarrow$ & $8.2$ \\
                $8$ & $6$ & $0.127$ & $0.035\pm 0.004$ & $72.44\downarrow$ & $11.3$ \\
                $16$ & $8$ & $0.224$ & $0.121\pm 0.013$ & $45.98\downarrow$ & $26.4$ \\
                $32$ & $10$ & $0.393$ & $0.246\pm 0.014$ & $37.40\downarrow$ & $87.9$ \\
                $64$ & $12$ & $0.457$ & $0.336\pm 0.029$ & $26.48\downarrow$ & $178.9$ \\ \hline
            \end{tabular}
        \egroup \\ (b) Optimization with $100$ generations. \\
    \end{tabular}
    \caption{Results averaged from three optimization routines of quantum noise parameters with our UNM model. Each routine is run for (a) $50$ and (b) $100$ generations of the genetic algorithm. HD (Pre) and HD (Post) are the HD between the probability distributions of the quantum computer and the simulator preoptimization and postoptimization (along with standard deviation, s.d., rounded up to three decimal points) respectively; $\uparrow$ or $\downarrow$ mean increase or decrease in the distance between the distributions. The size of workspace is the number of qubits necessary for the computation (i.e. ancilla included). The runtime showcased is the average of the three optimization routines. Note: since the HD (Pre) is the distance preoptimization, there are no multiple runs and hence, no need for standard deviation.}
    \label{table:optresults}
\end{table*}

In addition, we note that the efficiency of the parameters postoptimization declines with the size of the state space. The main reason for that is the fact that we kept the number of generations of the algorithm stable to $50$ iterations. We find that more generations of the GA during the experiments provide further improvement on the approximation of the quantum computer's distribution, even on quantum walks with larger state spaces (see Table \ref{table:optresults}(b)).

As a final remark we reiterate our decision not to include the relaxation and dephasing parameters in the optimization. From Table \ref{table:decoptresults} we see that the two-qubit quantum walk postoptimization offers a similar increase in efficiency for half the generations as when also optimizing the decoherence parameters (i.e., $84.85\%$ vs $85.48\%$ respectively). Similarly, for the three-qubit quantum walk and the same number of generations the optimization without the decoherence parameters approximates the quantum computer better than when the decoherence rates are optimized (i.e., $72.44\%$ vs $64.57\%$ respectively). See Appendix \ref{ap:decopt} for more details.

For the parameter optimization we used a MacBook Pro $2017$ computer with a $2.3$ GHz Intel Core i5 processor and $16$ GB of memory.

\subsection{Noise Parameters Analysis}\label{subsec:params}
Here, we compare the model parameters pre and postoptimization for the $N=4$ states case. We chose the smallest system as it has the smallest number of parameters optimized, but the same analysis can be carried out for a system of any size.

Table \ref{table:paramsanalysis} shows the relevant noise parameters for the $N=4$ state space quantum walk pre and postoptimization, after a single optimization routine. Overall we can say that our noise models operate closer to the computer for different error rates than the ones provided by the computer's calibrations. More specifically, comparing the parameters pre and postoptimization from Table \ref{table:paramsanalysis}, single-qubit operations and measurements on qubits $0$ and $1$ of the IBM quantum computer are noisier than the calibrations claim, with the opposite being true for qubits $2$ and $3$ and two-qubit operations on all qubit pairs.

\bgroup
\def\arraystretch{1.2}%
\setlength\tabcolsep{0.1cm}
\begin{table*}[!tb]
    \centering
    \begin{tabular}{c|c|c|c|c|c|c|c|c|c|c}
        \hline
        Optimization & Sq$(0)$ & Sq$(1)$ & Sq$(2)$ & Sq$(3)$ & CNOT$(0,1)$ & CNOT$(1,2)$ & CNOT$(2,3)$ & $M(0)$ & $M(1)$ & $M(2)$ \\ \hline \hline
        Pre & $0.000631$ & $0.000550$ & $0.000550$ & $0.000496$ & $0.015850$ & $0.011410$ & $0.021430$ & $0.036700$ & $0.080900$ & $0.032400$ \\
        Post & $0.000685$ & $0.000725$ & $0.000406$ & $0.000479$ & $0.010435$ & $0.010074$ & $0.013219$ & $0.038924$ & $0.090734$ & $0.039960$ \\ \hline
    \end{tabular}
    \caption{Pre and postoptimization noise parameters for $N=4$ state space system; Sq$(q)$ are the single-qubit gate error rates, including Hadamard and NOT gates, CNOT$(q,q^{\prime})$ are the two-qubit gate error rates and $M(q)$ are the measurement error rates for each qubit $q$ or pair of qubits $(q,q^{\prime})$, according to the architecture of the quantum computer.}
    \label{table:paramsanalysis}
\end{table*}
\egroup

The same analysis applied to the larger systems shows that our model performs closer to the computer when single-qubit operations and measurements are, in their majority, noisier than calibrated, whereas two-qubit operations tend to be less noisy. There are different factors that cause this. First of all, the length of the experiment. For larger experiments, where the computation is much longer than the times $T_{1}$ and $T_{2}$, it is very difficult to get a concrete conclusion through such an analysis. As shown above, for smaller computations (and not necessarily quantum walks), the above methodology could provide a very good picture of whether the quantum computer calibrations overestimate or underestimate each of the error rates. Secondly, the above findings are the averaged results of three optimization routines. This means that the claim of this analysis could still be an artefact of the randomness embedded within our parameter optimization technique. Further optimization runs for the same experiment could revoke this ambiguity. Unfortunately this endeavour could prove increasingly time consuming, especially for longer computations with a much larger number of noise parameters.

\section{Conclusions and Future Work}
In this paper we have presented an approach to modelling the noise in quantum computers that combines three sources of error, each modeled through quantum channels whose basic principles are well-known within the field. The model takes into account the architectural characteristics of the quantum computer (i.e., qubit connectivity) as well as various hardware-calibrated noise parameters in order to simulate the noisy quantum evolution within the computer. We have tested our unified noise model by evaluating its performance when executing quantum walks over different state spaces. Comparisons with the probability distributions from other simulated noise models, as well as the IBMQ $15$-qubit Melbourne machine, have shown that our unified model offers a better approximation of the quantum computer's noisy evolution. To further improve the efficiency of our noise model, we have implemented parameter optimization via a genetic algorithm. Experiments have shown that our optimized parameters offer a better approximation of the quantum computer behaviour that can be more than $84\%$ closer to the actual one.

A recent study \cite{Smart-2021} simulates the relaxation of stationary states in order to obtain spectroscopic fingerprints of their noisy. The results show that noise follows largely non-Markovian behaviour. They also suggest that quantum computers can be modelled as non-Markovian noise baths and analysed through simulations, thus providing interesting potential applications on error mitigation.

Within our work, the unified noise model employs a highly Markovian approach to simulate the noise and decoherence for all three quantum channels. As is evident by the results, this approach produces a satisfactory performance, especially on small quantum systems. It would be interesting to include non-Markovian noise into our framework and investigate its pros and cons.

Future additions to our unified model could include the consideration of further sources of error. Examples include, but are not limited to, errors from the Clifford group \cite{Gutierrez-2013}, or additional forms of decoherence, i.e., electromagnetic. Furthermore, within the context of this work, we consider SPAM errors to be purely errors of the hardware with the most common error to be a Pauli-$X$. It is possible that state preparation or measurement devices could cause other types of error, i.e., Pauli-$Z$ or -$Y$, or thermal decoherence or dephasing due to the duration of such operations. Our limitation of SPAM errors to exclusively Pauli-$X$ seems to work well-enough in terms of precision, especially postoptimization, but further improvements could be found considering such additional types of noise.

On the optimization side, the novelty of our work is twofold: first, the idea and framework for such a technique, has not been carried out before, to the best of our knowledge, and secondly, the unified noise model allows for such an optimization (unlike other models, i.e., the IBMQ models we compare with the UNM). Further improvements in the accuracy of our optimized parameters could be obtained by adjusting the characteristics of the genetic algorithm, like implementing a larger number of iterations during the optimization, or even experimenting with other optimization techniques.

Noise is one of the main challenges preventing universal and scalable quantum computation. Our work has shown that unifying noise sources on a single model results in a better approximation of the noisy evolution of a quantum computer. Additionally, hardware-calibrated noise parameters often produce simulations that deviate from the actual noise within the quantum computer. Our model is able to showcase this weakness, to present an insight on what the noise parameters look like within our unified model and to assist on limiting the gap between calibrated and simulated noise parameters. Finally, our approach to noise modelling can assist with the understanding of noise within quantum computers and consequently be utilized during the design or testing of error correcting methods or calibration techniques and attempts to minimize the noise in near-term quantum computers.

\section{Acknowledgements}
This work was supported by the Engineering and Physical Sciences Research Council, Centre for Doctoral Training in Cloud Computing for Big Data [grant number EP/L015358/1].


\appendix

\section{Thermal Relaxation and Dephasing Model}\label{ap:trm}
This section supplements the Section \ref{subsec:thermal} description of the thermal decoherence and dephasing channel on the general case, i.e., when we also take thermal excitation into account. Here we consider the temperature of the quantum processor to be $\Theta \geq 0$. Having defined in Section \ref{subsec:thermal} the time parameters $T_{1}$, $T_{2}$, $T_{g}$ and the probabilities $p_{T_{1}}$ and $p_{T_{2}}$, as well as equation \eqref{eq:esp} and the probability of reset as $p_{\text{reset}}=1-p_{T_{1}}$, we can identify the following forms of noise for the case where $T_{2}\leq T_{1}$:
\begin{itemize}
    \item Dephasing. A phase-flip which occurs with probability $p_{Z}=(1-p_{\text{reset}})(1-p_{T_{2}}p_{T_{1}}^{-1})/2$.
    
    \item Reset to $\ket{0}$. This represents a quantum decay, or a jump to the ground state, and occurs with probability $p_{\text{reset}_{0}}=(1-w_{e})p_{\text{reset}}$.
    
    \item Reset to $\ket{1}$. This represents a spontaneous excitation, or a jump to the excited state, and occurs with probability $p_{\text{reset}_{1}}=w_{e}p_{\text{reset}}$.
    
    \item Identity. In this case, nothing happens to the state, or otherwise, the identity, $I$, occurs with probability $p_{I}=1-p_{Z}-p_{\text{reset}_{0}}-p_{\text{reset}_{1}}$.
\end{itemize}
where $w_{e}$ is given as in equation \eqref{eq:esp}.

The operators for this case follow simply from the forms of noise described above as
\begin{equation}
    \begin{aligned}
        K_{I} &= \sqrt{p_{I}}I, \\
        K_{Z} &= \sqrt{p_{Z}}Z, \\
        K_{\text{reset}_{0}} &= \sqrt{p_{\text{reset}_{0}}} \ket{0}\bra{0}, \\
        K_{\text{reset}_{1}} &= \sqrt{p_{\text{reset}_{1}}} \ket{1}\bra{1}.
    \end{aligned}
\end{equation}
and the operator-sum representation describing the quantum channel will be
\begin{equation*}
    \rho \mapsto \mathcal{N}(\rho) = \sum_{k\in \left\{ I,Z,\text{reset}_{0},\text{reset}_{1} \right\}}K_{k}\rho K_{k}^{\dagger}.
\end{equation*}

If $2T_{1}\geq T_{2} > T_{1}$ then a Choi-matrix representation of the form of equation \eqref{eq:choi} is used, as in Section \ref{subsec:thermal}. In general, a Choi matrix is defined as
\begin{equation*}
    C = \sum_{i,j} \ket{i}\bra{j} \otimes \mathcal{E}\left( \ket{i}\bra{j} \right),
\end{equation*}
with $\mathcal{E}(\cdot)$ an arbitrary quantum channel. For a single-qubit case, we have $i,j=\{0,1\}$.

The transition from Choi-matrix representation to operator-sum representation can be done via the spectral theorem as
\begin{equation*}
    C = \sum_{j=1}^{r} v_{j}v_{j}^{\dagger}
\end{equation*}
for vectors $v_{1},\dots,v_{r}$ and $r=\text{rank}(C)$. We can then deduce the Kraus operators to be the operators $K_{1},\dots,K_{r}$ such that $\text{vec}(K_{j})=v_{j}$, for $j\in \{ 1, \dots, r \}$.

If the Choi matrix is Hermitian, then, given an isomorphism from $\mathbb{C}^{n^{2}}$ to $\mathbb{C}^{n\times n}$, the Kraus operators can be expressed as
\begin{equation*}
    K_{\lambda} = \sqrt{\lambda}\Phi(v_{\lambda}),
\end{equation*}
where $\lambda$ are the eigenvalues and $v_{\lambda}$ the eigenvectors of $C$.

If the Choi matrix is not Hermitian, or if its eigenvalues are negative, then singular value decomposition (SVD) is applied. Let the SVD of the Choi matrix be
\begin{equation*}
    C = U \Sigma V^{\dagger},
\end{equation*}
where $\Sigma = \text{diag}(\sigma_{1}, \dots, \sigma_{n})$, $\sigma_{i}\geq 0$, $U=(u_{1}|\dots|u_{n})$ the left singular vectors and $V=(v_{1}|\dots|v_{n})$ the right singular vectors. This leads to two sets of Kraus operators, one for the left and one for the right map, which can be expressed as
\begin{equation*}
    \begin{aligned}
        K_{i}^{l} &= \sqrt{\sigma_{i}}\Phi(u_{i}) \\
        K_{i}^{r} &= \sqrt{\sigma_{i}}\Phi(v_{i}).
    \end{aligned}
\end{equation*}
If the left and right Kraus operators are not equal, i.e. $u_{i} \neq v_{i}$ for some $i\in [1,\dots,n]$, then they do not represent a completely positive trace preserving map, triggering an error in the thermal relaxation model.

\bgroup
\def\arraystretch{1.2}%
\setlength\tabcolsep{0.32cm}
\begin{table*}[!t]
    \centering
    \begin{tabular}{c|c|c|c|c|c}
        \hline
        No. States ($N$) & Size of Workspace & HD (Pre) & HD (Post) & $\%$ Distance & CPU Runtime ($\times 10^{-3}$ sec) \\ \hline \hline
        $4$ & $4$ & $0.033$ & $0.0048$ & $85.48 \downarrow$ & $8.1$ \\
        $8$ & $6$ & $0.127$ & $0.045$ & $64.57 \downarrow$ & $11.9$ \\ \hline
    \end{tabular}
    \caption{Simple table following the trend of Table \ref{table:optresults}.}
    \label{table:decoptresults}
\end{table*}
\egroup

\section{Circuit for Discrete-time Quantum Walk}\label{ap:qwcirc}
\begin{figure}[!tb]
    \begin{center}
        \includegraphics[width=8cm]{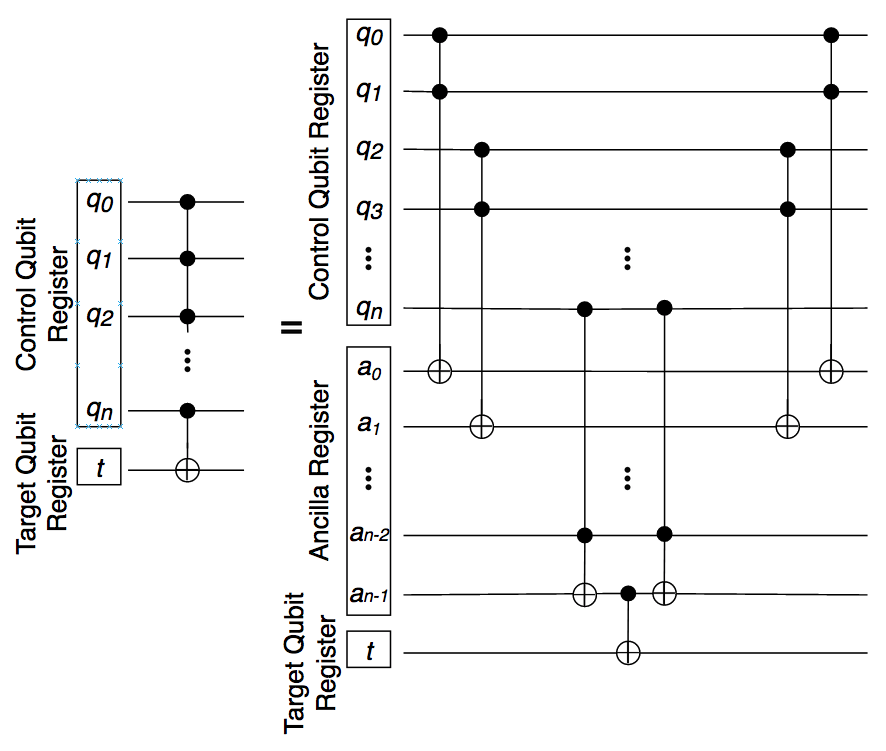}
    \end{center}
    \caption{Generalised CNOT gate with $n$ control qubits ($q_{0}$ to $q_{n}$), $n-1$ ancilla qubits ($anc_{0}$ to $anc_{n-1}$) and one target qubit ($tgt$).}
    \label{fig:GenTof}
\end{figure}

The approach used to implement the discrete-time quantum walk within this circuit was first introduced in \cite{Douglas-2009-EffWalk}. It uses what is called a generalized control quantum gate, i.e., a quantum gate controlled by two or more qubits. In \cite{KGeorgo-2020-rots} we showcased a general strategy to implement a generalized CNOT gate using the expansion of the form presented in Figure \ref{fig:GenTof}.

\begin{figure}[!tb]
    \begin{tabular}{c}
          \includegraphics[width=7cm]{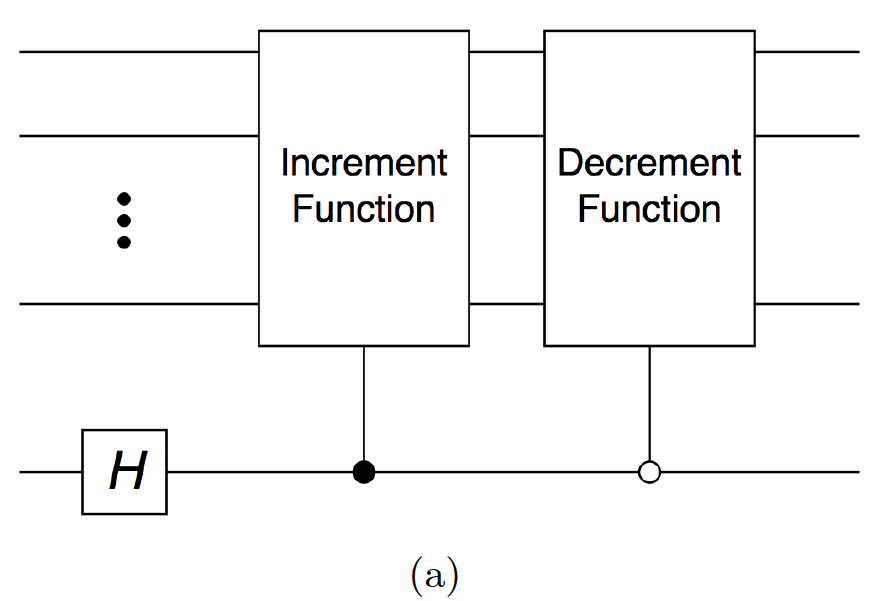} \\
          \includegraphics[width=8.5cm]{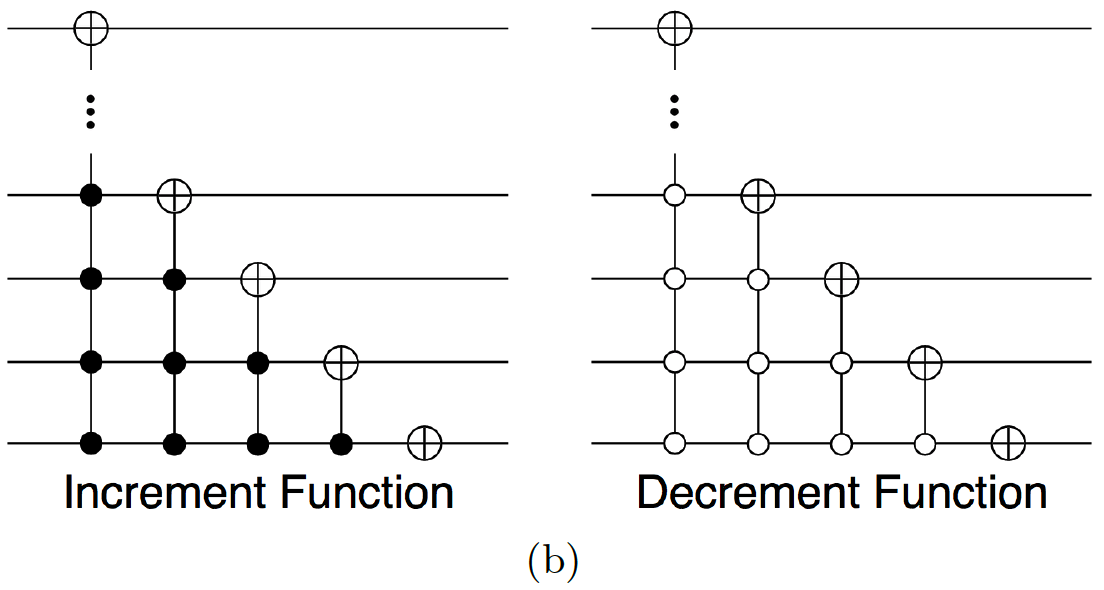} \\
    \end{tabular}
    \caption{(a) Implementation of one step for the quantum walk of a particle. (b) Quantum circuits for increment and decrement operations. A filled control circle means that the control qubits have to be in state $\ket{1}$ in order for the operation to occur. An empty control circle means they have to be in state $\ket{0}$.}
    \label{fig:incredecrcirc}
\end{figure}

The quantum walk circuit can be constructed as a sequence of two functions, an \textit{increment}, which essentially increases the state of the quantum register, and a \textit{decrement}, which decreases it. These two functions are implemented for an arbitrary number of qubits as shown Figure \ref{fig:incredecrcirc}.

\section{Genetic Algorithm Optimization Including the Decoherence Parameters}\label{ap:decopt}
Here we present the results of parameter optimization when the relaxation and dephasing rates $T_{1}(q)$ and $T_{2}(q)$ are included. These parameters need to be considered per qubit in the system, including the ancilla and coin qubits. Considering the two optimization routines analyzed in Section \ref{subsec:paramopt}, for a state space of $N=4$ the number of parameters that need optimizing are $17$: $9$ parameters that represent the hardware infidelities and SPAM errors, as shown in Section \ref{subsec:paramopt}, plus $8$ relaxation times $T_{1}(q)$ and $T_{2}(q)$, one for each of the four qubits in the workspace. For $N=8$ the number of parameters is $26$: $14$ for the hardware and SPAM errors plus $12$ relaxation and dephasing times, one for each of the six qubits in the workspace. Table \ref{table:decoptresults} shows the results of a GA parameter optimization routine with $100$ generations.

When comparing the results of this table to those showcased in Table \ref{table:optresults}(b) we can draw some very interesting conclusions. As is evident from the percentage of decrease in the distance between the distributions of the simulated quantum walk and the evolution of the quantum computer, the optimization performs better in both cases when the decoherence parameters $T_{1}(q)$ and $T_{2}(q)$ are excluded from the optimization. There are a couple of reasons for this, the most important of which is the fact that with more parameters to optimize, less space of the potential optimal parameters is searched. Secondly, the thermal relaxation and dephasing model requires $T_{2}(q)\leq 2T_{1}(q)$, which means that we have to enforce this condition within the parameter optimization, something that will, again, limit the space within which the GA can look for the optimal parameters. 

Furthermore, we observe an increase in the computational resources necessary for the optimization when these parameters are included. This derives from the aforementioned decrease in the performance of the optimization routine due to the increased number in parameters. In other words, to get the same increase in efficiency of our simulations we would need to run more generations of the genetic algorithm. This cripples the performance of the optimization routine, especially for the experiments with a very large workspace.

\bibliography{bibphy}{}

\end{document}